%Paper: gr-qc/9512044
%From: BELGIORNO@mi.infn.it
%Date: Thu, 28 Dec 1995 18:11:38 +0100 (WET)

\input phyzzx
\physrev

\REF\tho{G.`t Hooft\journal Nucl. Phys. &256B(1985)727.}
\REF\mann{R.B.Mann,L.Tarasov,A.Zelnikov\journal Class.
Quant. Grav. &9(92)1487}
\REF\bkls{L.Bombelli, R.K.Koul,
J.Lee and R.Sorkin\journal Phys. Rev. &D34(86)373.}
\REF\susug{L.Susskind and J.Uglum\journal Phys.Rev.
&D50(1994)2700.}
\REF\fur{D.V.Fursaev\journal Mod.Phys.Lett. &A10(1995)649.}
\REF\solo{S.N.Solodukhin\journal Phys.Rev. &D51(95)609, {\sl
ibid.} {\bf D51}(1995), 618.}
\REF\fursolo{D.V.Fursaev,S.N.Solodukhin-On One-Loop Renormalization of
Black Hole Entropy- Pre\-print hep-th/9412020.}
\REF\frono{V.Frolov and I.Novikov\journal Phys. Rev. &D48(93)4545.}
\REF\bfz{A.O.Barvinsky, V.P.Frolov and A.I.Zelnikov\journal
Phys.Rev &D51(1995)1741.}
\REF\frobo{V.Frolov and I.Novikov-Black Hole Physics-Kluwer
1992.}
\REF\dowken{J.S.Dowker and G.Kennedy\journal J.Phys.A. &11(1978)895.}
\REF\dosho{J.S.Dowker, J.P.Shofield\journal Jou. Math. Phys.
&31(90)808.}
\REF\froa{V.P.Frolov\journal Phys. Rev. Lett. &74(95)3319.}
\REF\cogno{G.Cognola, K.Kirsten and L.Vanzo\journal
Phys. Rev. &D49(94)1029.}
\REF\takume{Y.Takahashi and H.Umezawa\journal Collective
Phenomena &2(75)55.}
\REF\lafl{R.Laflamme\journal Nucl. Phys. &B324(89)233.}
\REF\jac{T.Jacobson\journal Phys. Rev. &D50(94)R6031.}
\REF\bellib{F.Belgiorno, S.Liberati (1995)-Divergences Problem in Brick-Wall
Model-Preprint Roma1/hep/95/n.1093. Accepted for publication in
{\sl Phys. Rev.} {\bf D}.}
\REF\page{D.N.Page\journal Phys. Rev. &D25(82)1499.}
\REF\pavon{D.Pav\'on and W.Israel\journal Gen.Rel.Grav.
&16(84)563.}
\REF\domi{J.S.Dowker\journal Class. Quantum Grav. &11(94)L55.}
\REF\tadtak{S.Tadaki,S.Takagi\journal Progr.Teor.Phys. Progr.
Lett. &73(85)288.}
\REF\belg{F. Belgiorno-Ph.D. Thesis-Universit\`a degli Studi di Milano (1995).}
\REF\haweu{S.W.Hawking-Euclidean Quantum Gravity, in {\sl Recent
Developments in Gravitation}-Cargese Lectures, eds. M.levy and
S.Deser,(Plenum 1978).}
\REF\hawcen{S.W.Hawking-The Path--Integral Approach to Quantum Gravity,
in {\sl General Relativity: An Einstein Centenary Survey}-
eds. S.W.Hawking and W.Israel (CUP, 1979).}
\REF\gross{D.J.Gross and M.J.Perry, L.G.Yaffe\journal Phys. Rev.
&D25(82)330.}
\REF\gibhaw{G.Gibbons and S.W.Hawking\journal Phys. Rev. &D15(77)2752.}
\REF\gihaw{G.Gibbons and S.W.Hawking\journal Phys. Rev. &D15(77)2738.}
\REF\work{In preparation.}
\REF\allen{B.Allen\journal Phys. Rev. &D33(86)3640.}
\REF\barbon{J.L.F.Barbon\journal Phys. Rev. &D50(94)2712.}
\REF\hawzet{S.W.Hawking\journal Commun. Math. Phys. &56(77)133.}
\REF\brotte{M.R.Brown and A.C.Ottewill\journal Phys. Rev.
&D31(85)2514.}
\REF\bardet{J.L.F.Barbon and R.Emparan (1995)-On Quantum Black
Hole Entropy And Newton Constant Renormalization-Preprint
PUPT-1529 EHU-FT 95/5 hep-th/9502155.}
\REF\zerva{G.Cognola,L.Vanzo and S.Zerbini (1995)-One-loop
Quantum Corrections to the Entropy for a 4-dimensional Eternal
Black Hole-Preprint UTF 342}
\REF\deal{S.P.de Alwis, N.H.Ohta (1994) -On the Entropy of Quantum Fields
in Black Hole Background-Preprint COLO-HEP-347, OU-HET
 207, hep-th/9412027.}
\REF\dealot{S.P.de Alwis, N.H.Ohta (1995) -Thermodynamics of
Quantum Fields in Black Hole Backgrounds-Preprint COLO-HEP-357, OU-HET
 209, hep-th/9504033.}
\REF\doshof{J.S.Dowker, J.P.Shofield\journal Nucl. Phys.
&B327(89)267.}
\REF\ross{S.W.Hawking, G.T.Horowitz and S.F.Ross (1994)-Entropy,
Area and Black Hole Pairs- Preprint NI-94-012, DAMPT/R 94-26,
UCSBTH-94-25, gr-qc/9409013.}
\REF\moretti{W.Moretti (1995)-Hessling's Quantum Equivalence Principle
and the Temperature of an Extremal Reissner-Nordstr\"{o}m Black
Hole-Preprint - UTF 363, gr-qc/9510016.}
\REF\bhe{F.Belgiorno, S.Liberati, M.Martellini-Work in progress.}
\REF\isr{W.Israel\journal Phys.Lett. &57A(1976)107.}

\hsize=14cm
\vsize=20cm
\voffset=1cm
\hoffset=5mm
\baselineskip=20pt

\def\rig{\rightarrow}

\def\beh{\beta_{h}}
\def\os{\beta=\beta_{h}}
\def\bep{{\beta}^{\prime}}

\def\pabe{\partial_{\beta}}

\def\pano{\par \noindent}
\def\sch{Schwarzschild}

\def\ee{entanglement entropy}

\def\text{Tr_{ext}}

\def\calh{{\cal H}}

\def\rh{r_{bh}}

\def\ricqua{R_{\mu\nu} R^{\mu\nu}}
\def\ricmn{R_{\mu\nu}}

\def\riequa{R_{\mu \nu \lambda \rho} R^{\mu \nu \lambda \rho}}

\def\dx{d^{4}x \sqrt{g}}
\def\oms{{\omega}^{\mu}}
\def\omg{{\omega}_{\mu}}
\def\om{\omega}
\def\ovg{{\overline{g}}_{\mu \nu}}
\def\gmn{g_{\mu \nu}}
\def\tmn{T_{\mu \nu}}
\def\thh{T^{HH}_{\mu \nu}}

\def\therm{T^{therm\ \nu}_{\mu}}

\def\overw{{\overline{W}}}
\def\dew{\Delta {W}}

\def\rb{R}

\def\rh{r_{bh}}

\def\lp{l_{pl}}
\def\sul{S_{1 loop}^{matter}}
\def\sho{S_{horizon}}
\def\svo{S_{volume}}

\pubnum={IFUM 521/FT}
\date={11/1995}

\titlepage
\title{Hawking Radiation Entropy and Horizon Divergences}

\author{F. Belgiorno\foot{Also Sezione I.N.F.N. di Milano,
 20133 Milano, Italy. E-mail: belgiorno@vaxmi.mi.infn.it}
}
\address{Dipartimento di Fisica, Universit\`a di Milano, 20133 Milano, Italy}

\author{M. Martellini\foot{On leave of absence from Dipartimento di Fisica,
Universit\`a di Milano, 20133 Milano, Italy and also Sezione I.N.F.N. di
Pavia, 27100 Pavia, Italy. E-mail: martellini@vaxmi.mi.infn.it}}

\address{The Niels Bohr Institute, University of Copenhagen,
DK-2100 Copenhagen $\phi$, Denmark}
\centerline{\sl and Landau Network at Centro Volta, Como}
\centerline{\sl 22100 Como, Italy}

\abstract{In this paper we review the problem of
divergences in one--loop thermodynamical quantities for matter
fields in thermal equilibrium on a black hole background. We discuss
a number of results obtained for various thermodynamical quantities.
Then we discuss the ansatz called ``literal
interpretation" of zeroth law of black hole mechanics and try to
explain the diseases of the conical defect procedure in light of
this ansatz. Finally, an analysis of the consequences implied by our
ansatz on the calculation of the partition function is made.}

\endpage

\chapter{Introduction}

In this work we will discuss the following problem: the
divergence of the 1 loop entropy for matter fields in equilibrium
with a \sch\ black hole.\pano
In fact, the underlying problem is the calculation of the 1 loop
matter fields partition function on the \sch\ background.\pano
The first calculations on this topic are due to G. `t Hooft
[\tho]: using a WKB approximation for the eigenvalues of a scalar field
hamiltonian on the Schwarz\-schild background, `t Hooft finds that
thermodynamical quantities as free energy, internal energy and
entropy have contributions divergent for the radial coordinate
$r\rig \rh$, where $\rh$ is the black hole radius.
So one must introduce a short--distance cut--off
$\epsilon$ representing a radial proper distance from the
horizon.
The divergences in the thermodynamical quantities behave as
 $\epsilon^{-2}$\rlap.\foot{The divergence
problem for generic value of the space--time dimension is
discussed in [\mann].}\pano
`t Hooft proposal to face with these divergences is the so called
brick wall model: his ansatz is to consider all the black hole
entropy $S_{bh}=A/4$, where $A$ is the horizon area,
as due to the radiation external to the black hole; then the
cut--off is to be fixed to the value $\epsilon \sim \lp$. This
means that the ``brick wall" proper distance from the horizon
coordinate $\rh$ is of the order of the Planck length; besides, the
brick wall cut--off is shown to be an universal property: it is
independent from the black hole mass.\par
An unsatisfactory result of this model, underlined by `t Hooft,
is that the leading internal energy contribution of the radiation
is a not negligible fraction of the black hole mass: it is about
half of the black hole mass. This fact represents a consistency
problem for the approximation of fixed black hole background and
negligible backreaction of the linearized theory.\pano
The most interesting hypothesis in the model is `t Hooft idea of
a link between the divergence problem and the unsolved problem of
the explanation of black hole entropy in terms of Statistical
Mechanics. The proportionality of the leading divergent term of
the matter entropy to the black hole area A is an indication that `t
Hooft ansatz could be right.\par
Similar results about proportionality to the area and horizon
divergences are successively found by Bombelli et al. [\bkls].
 They study what they call ``\ee": the entropy due to the
ignorance of an observer external to the horizon about the fields
correlations existing between points internal and external to the
horizon.
Recently, the divergence problem has been analyzed by many
authors. We will call ``entanglement program" the research
program identifying a strong link between
\item{a)}{the 1 loop horizon
divergence of the matter contribution to the entropy;}
\item{b)}{the problem of a statistical mechanical explanation of
black hole entropy.}
In this program `t Hooft ideas and the concept of ``\ee" are
often considered on the same foot, although they are not quite
equivalent.\par
The major topic in the entanglement program has been the entropy
divergence. Between the current proposals of solution for this
problem, we will take into account essentially the following two:
\item{1]}{to reabsorb (Susskind and Uglum [\susug])
the divergence in the gravitational coupling
constant $G$: studying the analogous divergence problem on the
Rindler background, Susskind and Uglum find that the same
quadratic divergence of the entropy enter into the
renormalization of the coupling $G$ in the effective action for
matter fields and gravitational field; the renormalization of $G$
implies the renormalization of the entropy. This approach has
been extensively pursued by Fursaev and Solodukhin
[\fur,\solo,\fursolo]; the latter author discovered a logarithmic
divergence in the entropy requiring to be renormalized
the introduction in the
gravitational action of terms quadratic in the curvature;}
\item{2]}{to find a physical mechanism
to implement the brick wall model. Frolov et al. [\frono,\bfz]
identify
this mechanism in the quantum fluctuations of the black hole
horizon. The spreading of the radial coordinate $\rh$ is of the
order of the Planck length [\frono,\frobo] (when expressed
in radial proper distance from the horizon). The quantum
fluctuation of the black hole radius acts as a physical thickness
of the right size: used as a cut--off, it gives an horizon
contribution for the matter entropy of the order of the black hole
entropy.}\pano

We will present the main ideas and results underlying it
and then we will explain our proposal. The
plan of our work consists of the following paragraphs: in 2 we
resume the divergence problem in its various aspects; in 3
a further discussion of the divergence problem is made in
light of the discrepancy between the finiteness of the internal
energy calculated by mean of the Hartle--Hawking tensor and
its divergence in conical calculations; in 4 we will propose
the "intrinsic thermodynamics approach"; it is essentially a literal
interpretation of the zeroth law of black hole mechanics and
it is a candidate to solve the divergence problem. It represent
an possibly coherent ansatz that allows us to include in a systematic
picture various known results about Hawking radiation.
In 5 we will analyze which consequences on the calculation of the partition
function our main proposal have. Last paragraph involves a further
discussion.

\chapter{Divergence problems}

In this paragraph we review the problem of the 1-loop divergence of
the matter entropy contribution to the black hole entropy.\par

Let us consider preliminarily a mathematical problem: the
1-loop scalar field entropy calculation on the Rindler Wedge (RW).\par
\noindent The RW represents the limit of infinite mass of the
\sch\ space-time. The leading term of the 1-loop entropy is
$$
S_{horizon}={A \over {48 \pi}} {1 \over {\epsilon^2}}
\eqn\rin
$$
where $\epsilon\equiv {\rm proper\ distance\ from\ the\ horizon}$;
$\sho$ is a term diverging on the horizon.\par \noindent
If the horizon is kept fixed and if one does not try to implement
the renormalization approach,
the divergence is not avoidable: it could be
interpreted, in the standard statistical mechanics view, as a progressive
growing dense [\susug]\
 of the states of the Hilbert space available for the scalar
field as far as one approach the horizon; on the horizon, they become
infinite.\par  \noindent
We begin our discussion about the black hole entropy divergence
by revisiting `t Hooft brick-wall model [\tho]; we limit
our considerations to a spin-zero massless scalar field in equilibrium
with the black hole. The quantum field theory fluctuations
at the Hawking temperature give the following leading 1-loop contributions
to the entropy [\tho]:
$$\eqalign{
\sul &=\sho + \svo\cr
\sho &={{8 {\pi}^{3}} \over {45}}\ {{(2M)^4} \over {(\beta)^3}} {1 \over
{h}}\cr
\svo &={{8 {\pi}^{3}} \over {135}}\ {V \over {(\beta)^3}}\cr}
\eqn\thofff
$$
where $V$ is the box volume, $\beta$ is the inverse temperature
 and $h$ is a cut-off
$$
h\equiv {\rm Inf}(r-2M)
\eqn\off
$$
introduced in the calculation to avoid the divergence for
$r \rightarrow 2M$: a strong growing dense of the states appears also
near the horizon
of the \sch\ space-time.\par \noindent
 If we take {\sl ad hoc}
$$
h={1 \over {720 \pi M}}  \equiv h_{bw}
\eqn\hbw
$$
the horizon contribution is exactly the Bekenstein-Hawking value
$S_{bh}=4 \pi M^{2}$; the interpretation of this ad hoc position
is that the black hole
entropy is entirely an {\sl external} property, related with a hot
brick-wall at a planckian proper distance from the horizon.\par
However, the entropy is not the only thermodynamical quantity
divergent on the horizon: indeed, `t Hooft results for the free
energy and the internal energy contributions near the horizon are
$$\eqalign{
F_{horizon}&=-{{2 {\pi}^{3}} \over {45}}\ {{(2M)^4} \over
{(\beta)^4}} {1 \over {h}}\cr
E_{horizon}&={{2 {\pi}^{3}} \over {15}}\ {{(2M)^4} \over
{(\beta)^4}} {1 \over {h}}.\cr}
\eqn\thofren
$$
In [\tho]
is also underlined the relevance of the neglected backreaction of
the radiation mass: indeed the horizon contribution of the
radiation internal energy in \thofren\
 evaluated for $h=h_{bw}$ and for $\os$ is $3/8 M$. For the free
energy one gets the value $-1/8 M$.\par
In [\frono,\bfz] the brick-wall cut-off arises physically from the
quantum fluctuations of the horizon.
This position allows to implement a dynamical explanation
of the black hole entropy: as in the brick-wall model,
the leading term represents not a new term
to be summed to the black hole entropy, but again
the black hole entropy itself.
It is identified with the so called ``dynamical entropy" [\froa].
The relevance of the quantum trembling of the horizon is obvious:
it avoids the above infinite growing dense of the states; the same role
plays also $h_{bw}$ in the brick-wall model.\par
\noindent Introducing the proper distance cut-off
$$
\epsilon\sim 2 \sqrt{\rh h}
\eqn\prd
$$
we can write the divergent terms as it follows:
$$\eqalign{
\sho&={A \over {360 \pi}} {1 \over {\epsilon^2}} +
{1 \over {90 \pi}} log({1 \over {\epsilon^2}}) \cr
&=c_{1}\ {A \over{4}} {1 \over {\epsilon^2}} +
c_{2}\ log({{L^{2}} \over {\epsilon^2}});\cr}
\eqn\geo
$$
$A$ is the horizon area, $L$ is an infrared cut-off.\par \noindent
In \geo\ it appears also the logarithmically divergent term,
firstly found by Solodukhin [\solo].\par

\vfill
\eject

\section{QFT in Heisenberg Representation}

In finite temperature quantum field theory on a {\sl
fixed}\foot{As in `t Hooft pioneering calculations.}
static curved background\foot{
The path integral
functional measure
$$\eqalign{
D[\Phi]&=\prod_{x} d\phi (x) g^{1\over{4}} (x)\cr
g&=det \gmn\cr}
\eqn\funme
$$
is the usual covariant one.} there are essentially two
schemes for calculating the thermodynamical quantities: the
Dowker-Kennedy [\dowken] scheme and the Fursaev-Solodukhin
[\fursolo] one.\par
The first is limited to conformally coupled scalar fields and it is
based on the trick of passing to the optical manifold
$$
\ovg={\gmn \over{g_{0 0}}}.
\eqn\opt
$$
in order to
implement the Matsubara sums on the periodic euclidean time and
taking into account the difference w.r.t. the original static
manifold by mean of a functional jacobian:
$$\eqalign{
W[\gmn]&=W[\ovg]+\Delta W[\gmn, \omega]\cr
&\equiv \overw+\Delta W\cr
\ovg&= e^{-2 \omega} \gmn.\cr}
\eqn\dows
$$
For a manifold without boundary the jacobian is
$$\eqalign{
\Delta W[\gmn, \omega]&={1\over{2880 {\pi}^2}}
\int \dx [(\riequa-\ricqua+\nabla^{2} R) \omega\cr
& -2 \ricmn \oms \oms - 4
\oms \omg \nabla^{2} \om + 2 (\oms \omg)^{2}+3 (\nabla^{2} \om)^{2}]\cr
\omg & \equiv \nabla_{\mu} \omega.\cr}
\eqn\delw
$$
$\overw$ is the optical manifold contribution.
The boundary conditions are the
Dirichelet ones. For further details see
[\dowken,\dosho].
Given the relation between free energy and effective action
$$
\beta F[\beta]=W
\eqn\eliw
$$
we get
$$
F[\beta,\gmn]=F[\beta, \ovg]+\Delta F[\gmn, \om].
\eqn\free
$$
In \free\ the second term does not depend on $\beta$: the
$\beta$-dependence is entirely in the optical term
$$
F[\beta, \ovg]=-{{\pi}^2 \over{90}} {{\overline{c}_{0}} \over{{\beta}^4}}
+\ldots
\eqn\fhke
$$
Free energy and entropy are
$$\eqalign{
E[\beta, \gmn]&=\pabe (\beta F[\beta, \ovg])+\Delta F[\gmn, \om]\cr
S[\beta, \gmn]&=(\beta \pabe-1) (\beta F[\beta,\ovg]).\cr}
\eqn\enen
$$
The term $\Delta F$ in \enen\ for the internal energy
is usually omitted in the high temperature expansions, whose
leading terms are characterized by negative powers in $\beta$.\par

\par
The second one is based on an heat kernel expansion near the
horizon, taking into account the peculiar nature of the manifold.
It is in this case possible to implement the heat kernel periodicity
in the euclidean time without passing to the optical
manifold\rlap.\foot{The euclidean time is the angular variable
of the cone and it is possible to obtain an heat kernel on the
cone using periodic boundary conditions for the angular variable:
see [\fur] and references therein.}\par

In the second scheme it is to be noted that at the Hawking temperature
the free energy is finite:
it does not suffer of the divergence problem on the horizon characteristic of
the entropy and of the internal energy. The claim of free energy
finiteness appears in
[\froa,\fur] (for the Rindler case it is evident from the inspection of formula
(31) in ref. [\cogno]). \par \noindent
The structure of the divergence for $\beta
\neq \beta_{h}$ is very clear in [\fur, \solo]: the effective action has a
surface divergent term multiplied by a coefficient $\beta$-dependent
having a simple zero for $\beta=\beta_{h}$ (in the following we will call
``on shell" the thermodynamical quantities evaluated at $\beta=\beta_{h}$).
As a consequence, we cannot have finite $\beta$-derivatives of
the free energy, even if evaluated on shell.\par
In both these schemes,
not only the entropy but also the internal energy is divergent on the horizon.
Indeed, the leading optical contribution to the
internal energy in the Dowker-Kennedy scheme
gives
$$
E_{rad}={{\pi}^{2} \over {30}} {1 \over {(\beh)^4}}\
\int\ d^{3}x\ \sqrt{g} {r^{2} \over {(r-2 M)^{2}}}.
\eqn\eint
$$
We find the same divergence as in the entropy. Using as a cut-off the quantum
trembling of the horizon or the brick-wall cut-off we get as leading term
$$
E_{horizon}={3 \over {8}} M
\eqn\thof
$$
the same value than in [\tho].
\par
In Fursaev-Solodukhin [\fursolo] renormalization approach, if we choose to
renormalize the matter entropy leading term in such a way to get
the black hole entropy value, we get straightforwardly for the
leading term of the internal energy divergent at the horizon the
following renormalized value
$$
E_{horizon}={M \over {2}};
\eqn\efur
$$
this is a value very similar to the one obtained by `t Hooft and
in \thof.\pano
The internal energy value obtained in both the approach is a very consistent
fraction of the black hole mass. So, as noted firstly by 't
Hooft [\tho], there is a strong backreaction effect.\par
Analogous results characterize the approach
of Barvinsky, Frolov, and Zelnikov's work [\bfz],
quoted in the following as BFZ; it represents an implementation
in Schroe\-dinger representation of the idea of
entropy as \ee\ and in a fixed background ansatz it is analogous
to Thermofield Dynamics\foot{See also
appendix 2, where for completeness
a brief summary of the formalism is given.} [\takume,\lafl,\jac]
that represents a natural framework in which thermodynamical entropy
is equivalent to an entanglement entropy.
For the entropy [\bfz], for the free
energy and the internal energy [\bellib] one can get the same
results than in `t Hooft brick wall model.\par
Also the other physical quantities deduced by mean of
$\beta$--derivatives are divergent on the horizon:
in [\bellib] the value
for the divergent specific heat is calculated in the framework of
the brick wall regularization and its problematic value
discussed; an analysis about interpretative problems concerning
the standard position
$$
S_{bh}\equiv S_{radiation,\ leading}
\eqn\bwa
$$
and its consequences is also carried in [\bellib].

\vfill
\eject

\chapter{Internal Energy and Hartle-Hawking Tensor}

There is another interesting question, related to the previous
discussion. Given the finite temperature renormalized action
$W$, the tensor
$$
\tmn={2 \over{\sqrt{g}}} {{\delta W} \over {\delta g^{{\mu}
{\nu}}}}
\eqn\tenact
$$
for $\os$ should be the Hartle-Hawking one:
$$
\tmn |_{\os}=\thh.
\eqn\tehh
$$
The internal energy relative to the quantum field in thermal equilibrium
can be calculated either by mean of
$$
E=-\pabe log Z
\eqn\efuz
$$
or by mean of
$$\eqalign{
E_{rad}&=\int d^{3}x \sqrt{g} \rho\cr
\rho&=-T_{0}^{0}\cr}
\eqn\ete
$$
where $\tmn$ is the quantum stress energy tensor
for the state of thermal equilibrium.
The internal energy calculated by mean of $\thh$ (Page
approximation [\page])
is finite on the horizon and it is given by [\pavon]
$$\eqalign{
E_{rad}&={1\over{1920 \pi}}{1\over{\rb}} y K(y)\cr
K(y)&={1\over{3}}y^{3}+y^{2}+3 y+4 log y -{22\over{3}}-5 y^{-1}-3
y^{-2}+11 y^{-3}\cr
y&\equiv {\rb \over{2 M}}.\cr}
\eqn\pavisr
$$
where a finite box radius $\rb$ is imposed.
As previously shown, the conical strategy gives a divergent
result that requires a further renormalization w.r.t. the
renormalization of the effective action or the introduction of
the brick-wall cut-off:
$$
E_{ren}\sim  {3\over{8}} M
\eqn\eren
$$
We find a situation in which two a priori equivalent patterns for
calculating the internal energy give rise to two different
results. Particularly, for the second way the internal energy
wants an ad hoc renormalization besides the free energy one.
Moreover, given the finiteness of \pavisr\ and of the effective action
for $\os$, then also the entropy
$$
S[\beta]|_{\os}=(\beta(E[\beta]-F[\beta]))|_{\os}
\eqn\ters
$$
should be finite. Thermodynamics seems to be really different
according to different (but a priori equivalent) schemes of
calculation.\par

We can understand the cause of the above problem as it follows.
The purely thermal part of the stress energy tensor is
$$
\therm={{\pi^{2}}\over{90}} {1\over{{\beta}^{4}}}
{1\over{(1-{{2 M}\over{r}})^{2}}} (\delta_{\mu}^{\nu}-4
\delta_{\mu}^{0} \delta_{0}^{\nu});
\eqn\tenther
$$
if one performs the integral in \ete, one gets that the resulting internal
energy is divergent and gives exactly the same result than
the brick-wall model.\pano
Besides, the following naive pattern (cf. Dowker  [\domi]) is
instructive: given the thermal stress energy tensor ``off
shell", i.e. for $\beta\neq \beh$ and given the internal energy
$$
E[\beta]=\int d^{3}x \sqrt{g} (-T_{0}^{0}[\beta])
\eqn\dowp
$$
one can obtain by integration in beta the free energy and the entropy:
$$\eqalign{
F[\beta]&={1\over{\beta}} \int^{\beta} d\bep
E[\bep]+{C\over{\beta}}\cr
S[\beta]&=\beta (E[\beta]-F[\beta])\cr}
\eqn\intdow
$$
where $C$ is an arbitrary constant. The brick wall results can
be obtained in this pattern putting $\os$ , $C=0$ and choosing the
brick-wall cut-off at the end of the calculation in \intdow.\pano
The suggestion coming from the above results is that the conical
strategy gives rise to the same disease one can obtain
in actually treating the Hawking radiation as a purely thermal
perfect gas. Due to vacuum polarization effects,
Hawking radiation is not, at least near the
horizon [\hawzet, \pavon], a purely thermal perfect gas.
One can verify that a phenomenological study [\tadtak] of
the Hawking radiation by mean of a purely thermal perfect gas approximation
suffers of the same divergences than the brick-wall model.

\chapter{The Relevance of an Intrinsic Thermodynamics}

We propose a possibility to overcome the problems arising in the
usual approach to the problem. We here further develop what we
called ``literal interpretation" of the zeroth law of black hole
mechanics. The main ideas were presented in [\belg] and sketched in
[\bellib].\pano
Hawking effect allows to identify the surface gravity with the
physical temperature, as measured by a static observer at
infinity, associated with a stationary black hole: black hole are
sources of thermal particles with
$$
T={k\over{2 \pi}}.
\eqn\surf
$$
A free variation of the temperature, or equivalently, of its inverse
$\beta$, as it is true in the canonical ensemble, implies that
the geometrical parameters entering in the surface gravity
are functions of the temperature by mean of \surf.
The equation defining the proper period of the
manifold
$$
\beta={{2 \pi}\over{k}}
\eqn\ans
$$
is interpreted as a constraint equation for the geometrical
parameters appearing in \ans; if the manifold is characterized by
$n$ geometrical scales $L_{j}$, then the equation
\ans\ becomes
$$
\beta={{2 \pi} \over{k (L_{1},\ldots,L_{n})}}.
\eqn\benn
$$
The above match gives rise to a $\beta$-dependence of the
metric which represents the main difference w.r.t. the standard case:
in standard manifolds, i.e. in manifolds not affected by Hawking
effect, equilibrium thermodynamics of matter fields is usually
calculated by mean of the canonical ensemble, in which the
inverse temperature $\beta$ is an external parameter
characterizing the equilibrium distribution of field
microstates and there is no link between temperature and geometry.\pano
Instead in a black hole manifold, due to Hawking effect, thermal
equilibrium is implemented only when matter fields and the black
hole are at the same temperature given by \ans.\pano
In the case of the \sch\ solution,
that is the classical black hole solution we mainly are interested in,
the relation $\beta={{2 \pi} \over {k}}=8 \pi M$
for the unperturbed \sch\ black-hole implies
$$
M={\beta \over{8 \pi}}.
\eqn\bel
$$
The relation \bel\ between its period in
euclidean time and the black hole mass is coherent with the
perturbative expansion of the path integral calculation of the
partition function: in general, the tree level value of \surf\
represents a link between geometry and thermodynamics that is not
modified by 1--loop matter field contributions unless one does take
into account their backreaction on the geometry.\par
Hawking's considerations [\haweu,\hawcen], Frolov's
ideas in [\froa] and in particular Gross et al. [\gross] statements
support the literal interpretation of \surf.\par

We get the following immediate consequences of the ansatz:\pano

\item{a)}{the horizon divergences are absent; they are
due to a unphysical shift from the physical temperature,
as it follows from Fursaev and Solodukhin heat kernel expansion;}
\item{b)}{quantum field theory at finite temperature in a fixed
black hole manifold becomes a theory in a
background field (the metric) depending on the tree
level temperature.
An expansion of the action around the background fields up to quadratic
terms gives [\gibhaw]
$$\eqalign{
S[\gmn, \Phi]&=S[g^{0}_{\mu \nu}+h_{\mu \nu}, \Phi^{0}+\phi]\cr
&=S[g^{0}_{\mu \nu},\Phi^{0}]+S_{2}[g^{0}_{\mu
\nu},\Phi^{0};\phi]\cr
&+S_{2}[g^{0}_{\mu \nu},\Phi^{0}; h_{\mu \nu}];\cr}
\eqn\expact
$$
we choose for simplicity $\Phi^{0}=0$.
The classical background term corresponds to the gravitational term,
that is to the Einstein--Hilbert action. The second quadratic term
in \expact\ depends only on the metric.
In the perturbative path--integral calculation of
the partition function [\gibhaw],
taking into account the $\beta$--dependence of the
background metric (in the \sch\ case cf. \bel)
it follows for the 1--loop matter field partition function
$$
log Z_{matter}\sim log
\int_{P(\beta)} [D\phi] e^{-S_{2}[g^{0}_{\mu \nu}(\beta);\phi]}.
\eqn\bacbe
$$
\pano
An analogous formula holds (cf. statements in [\gross]) for the 1--loop
gravitational contribution.
We stress again that the matter field backreaction on the
geometry at this level is not taken into account. So the
temperature has to be the one associated to the unperturbed
background manifold;}\par
\item{c)}{the matter field Hamiltonian, being a function of
the metric, becomes $\beta$--dependent:
$$
H=H[\Phi,\gmn (\beta)]=H(\beta).
\eqn\hamilbe
$$
The meaning of \hamilbe\ in relation with the canonical ensemble
will be discussed in the next section.}\pano
Although our discussion is here limited to black hole manifold,
the zeroth law ansatz is to be considered an approach valid
in every manifold characterized by an Hawking effect and can
straightforwardly generalized to a literal interpretation of the zeroth law of
horizon's thermodynamics [\gihaw], including the case of De Sitter spacetime.

\section{$H(\beta)$ and Canonical Ensemble}

In general in the canonical ensemble the following formulas are
valid:
$$\eqalign{
Z&=Tr(e^{-\beta H})\cr
\rho&={{e^{-\beta H}}\over{Z}}\cr
<O>&=Tr(\rho O)\cr}
\eqn\cano
$$
where $\rho$ is the density
matrix and $O$ is an observable operator. In \cano\ $\beta$
plays the double role of Lagrange multiplier and of physical
equilibrium temperature furnished to the system by a thermostat. The well
known relations
$$\eqalign{
E&=<H>=-\pabe logZ\cr
S&=-(\beta \pabe -1) logZ\cr}
\eqn\canonic
$$
and other similar relations allowing to calculate thermodynamical
quantities by mean of $\beta$--derivatives of the partition
function understand that the hamiltonian is not explicitly
temperature dependent, that is
$$
\pabe H=0.
\eqn\ordi
$$
If \ordi\ is not true, than the machinery of $\beta$--derivatives
generates spurious terms: e.g. one obtains
$$
-\pabe logZ=E+Tr(\rho \beta \pabe H(\beta)).
\eqn\enebe
$$
So, if the external parameter $\beta$ appears
explicitly in $H$, as it should be for fields in equilibrium
with a \sch\ black hole
\foot{We mean to come back on this topic [\work] for general
stationary black hole geometries and to
face with stringy black holes in a future work.}
in our ansatz, then thermodynamical quantities of interest
cannot be calculated by mean of $\beta$--derivatives unless
one finds an algorithm to get rid of spurious terms\foot{Note that \intdow\
gives results that are equivalent to the standard canonical
$\beta$-derivatives of the partition function only for a perfect
gas in a ``cool" geometry.}.\par
In the case of the Hawking radiation on the \sch\ background, a
conical strategy could seem a possible
solution to the problem: one could just make the following shift
$$
Z=Tr(e^{-\gamma H(\beta)})
\eqn\conicz
$$
where $\gamma$ now is to be considered a Lagrange multiplier
different from $\beta$ appearing in $H$. \conicz\ should allow to
take $\gamma$--derivatives without generating spurious terms, and
at the end of the calculation one could pose $\gamma=\beta$. In
fact the strategy does not work: it does not reproduce
the real physics of Hawking radiation, as seen.
The reason we can guess for this fact
is that \conicz\ implies to relax
the second property of the Lagrange multiplier $\gamma$ to be a
physical equilibrium temperature, and it could be assumed as a
posteriori proof corroborating our ansatz\rlap.\foot{A similar
conclusion can be deduced for Thermofield Dynamics: see appendix
2.}\par

As a possible way to overcome the above problems we propose
the following one:
let us calculate perturbatively the partition function according
to the standard finite temperature quantum field theory
$$
Z=Tr(e^{-\beta H(\beta)})=e^{-\beta F};
\eqn\staft
$$
as for the gravitational tree level, the assumption is
that the partition function is still related with \staft.
Being standard formulas \canonic\
inhibited, our choice is to calculate the internal energy
according to a general relativistic formula and to get the
entropy as for the tree level by mean of a thermodynamic relation:
$$\eqalign{
E&=\int d^{3}x \sqrt{g} (-T^{0}_{0})\cr
S&=\beta (E-F);\cr}
\eqn\solu
$$
the stress--energy tensor
appearing in \solu\ is the one for the Hartle--Hawking state.\pano
All the calculations are made ``on shell" and no $\beta$
derivative is involved.\par
Given the finiteness on the horizon of the effective action and
of the internal energy also the matter field entropy is
finite, because of the second formula in \solu.\pano
A remark is found in [\froa] about the finiteness of what the author calls
the ``thermodynamical" entropy. The main difference w.r.t.
[\froa] is that we get a finite result for the statistical
mechanical entropy of the radiation by mean of \solu, whereas
in [\froa] one faces with a infinite statistical mechanical
entropy to be compared with a finite thermodynamical
entropy\rlap.\foot{For more details see [\froa].}

\chapter{Matter Field Partition Function in \sch\ Manifold}

What follows it appears to be relevant to the discussion of
the calculation of thermal equilibrium quantities when the Hawking effect is
involved in the physics one is treating. The $\beta$--dependence of the
geometry modifies strongly the standard picture. To see why
we first consider the one--loop finite temperature
contribution for a scalar field on a fixed
curved background. The standard periodicity condition in the
imaginary time $\tau$ with period $\beta$ can be implemented by
mean of the following imagine sum on a non--periodic
(``zero--temperature") heat kernel $K_{\infty}$:
$$
K_{\beta}(x,y;s)=\sum_{n=-\infty}^{+\infty} K_{\infty}(x,y-n l
\beta; s)
\eqn\kabe
$$
where $s$ is the usual ``fifth coordinate" and $l$ is a four
vector in the same direction as the periodic coordinate.
The $n=0$ term in \kabe\ is ordinarily a zero
temperature--term [\dowken] and it is the only divergent one:
$$
K_{\beta}(x,y;s)=K_{\infty}(x,y;s)+\sum_{n \neq 0} K_{\infty}(x,y-n l
\beta; s).
\eqn\kabb
$$
If the zeroth law ansatz is right, then also the $n=0$ term is
$\beta$--dependent, because of the $\beta$--dependence of the geometry.
In the calculation of the partition function
$$
Z(\beta)=\int_{P(\beta)} [D\Phi] e^{-S[\Phi]},
\eqn\zeqq
$$
where $P(\beta)$ stays for the periodicity condition in
imaginary time (with period $\beta$) required for a boson
field, one cannot get rid of the $n=0$ term in thermodynamical
considerations.
Note that
$$
logZ={1\over{2}} \int_{0}^{\infty} ds {1\over{s}} \int dx
K_{\beta}(x,x;s).
\eqn\heke
$$
A way to compute the effective action is the one discussed in sec. 2.1:
in order to factorize the heat kernel in a part dependent only on the
euclidean time coordinate and in a part dependent on the spatial
section, a conformal transformation to the optical manifold is made.
For conformally coupled scalar fields in a
static manifold it is known (see e.g. [\brotte,\bardet,\zerva,
\deal,\dealot]) that the effective action for the static manifold is
given by the effective action $W$ in the related optical manifold
{\sl plus} a functional jacobian $\dew$ (sometimes called
``Liouville action" [\bardet,\dealot]) representing the
difference between trace anomaly contributions in the two
manifolds:
$$
logZ^{conf}_{static}=logZ^{conf}_{ultrastatic}+logZ_{jacobian}.
\eqn\logz
$$
The generalization to the nonconformal
coupling should be straightforward, and it consists simply in
writing
$$\eqalign{
logZ_{static}&=
logZ_{ultrastatic}+logZ_{jacobian}\cr
&+logZ_{non\ conformal\ jacobian}.\cr}
\eqn\janc
$$
where the last term is due to the nonconformally invariant
action terms (see [\doshof]).\par
Note that in the case of ``standard" thermodynamics (i.e. thermodynamics
not involving Hawking effect) the jacobian
contributions, being linear in $\beta$,
don't affect the thermodynamics involved in the
problem. Instead, in the case of the Hawking effect,
the jacobians are not simply proportional to $\beta$, and so they cannot
be a priori neglected in thermodynamical
considerations\rlap.\foot{In fact, in this case there are
horizon divergences due to the explosion
of the optical volume: the covariant volume density of the
optical manifold is divergent on the horizon; but also the
conformal jacobian is divergent. If, as it is expected, the
effective action is finite on the horizon, then the horizon divergent
contribution of the jacobian has to be canceled by the horizon divergences of
the optical effective action.}
This is also the reason why one can expect quite different results
by the calculation of the partition function by mean of
$$
log Z_{optical}(\beta)=-\sum_{k}log(1-e^{-\beta E_{k}})
\eqn\zetopti
$$
where $k$ is the index for single particle states, calculated in the
optical manifold, whose energy is $E_{k}$. The point is that
Allen's theorem [\allen] allowing the thermodynamical equivalence between
``thermal partition function" \zetopti\ and ``quantum partition function"
\zeqq\ does not hold, due to the $\beta$--dependence of the geometry.
We will discuss further this topic in appendix 1.

\section{Thermofield Dynamics}

The question that should be raised is: which is the place of
Thermofield Dynamics in the above picture? \pano
In Laflamme [\lafl] and in BFZ euclidean path integral
implementation of Thermofield Dynamics
(however remember BFZ remarks about the
difference of their approach w.r.t. the thermofield approach of
Laflamme [\lafl]) there lacks an explicit definition of the functional
measure for the Euclidean Path--Integral; Barbon and Emparan [\bardet]
underlined that Thermofield Dynamics results for the vacuum density
matrix are compatible only with the choice of the non--covariant
measure
$$
D[\Phi]=\prod_{x} d\phi (x) ({{\sqrt{g}}\over{g_{00}}})^{1\over{2}} (x).
\eqn\funmenon
$$
The measure in \funmenon\ is inferable by inspection in BFZ
of the scalar product: the eigenfunctions are
orthonormalized according to the scalar product of the optical manifold.
But it is possible also to choose the
covariant measure in the implementation of Thermofield Dynamics:
indeed, BFZ and Laflamme results can be related to a covariant measure
by understanding that a conformal transformation of the
metric to the optical manifold is made in order to consider
the scalar product in the optical manifold.
Then one gets the same results than in Laflamme and in BFZ, with
the only difference that one has to take into account also the above jacobian
factors due to the transformation to the optical manifold.\pano
We stress again that it is not a
substantially irrelevant problem the presence or the absence of
such jacobian factors; this discussion on Thermofield
Dynamics is made in order to show that, with the appropriate choice of
the functional measure, Thermofield
Dynamics can give the same results than the
standard finite temperature quantum field theory also in presence
of the Hawking effect.\pano

\chapter{Conclusions and Summary}

In order to underline the the main points of this work,
let us compare again the thermodynamics of a
massless quantum field in thermal equilibrium at the temperature
$\beta$ in a static manifold without horizons with the case of
thermal radiation in equilibrium with a black hole.\pano
In the first case, thermal and quantum partition
function give the same thermodynamics (terms in $logZ$
proportional to $\beta$ are in fact contributions to the
cosmological constant [\allen], shifting by a constant the value
of the free energy and of the internal energy, but leaving
unaffected the entropy).\par
In the black hole case, if we take $\beta$ as a free parameter,
and if we want to describe thermal equilibrium between Hawking
radiation and the black hole,
we have to give a $\beta$ dependence to the geometrical parameters involved
in the expression of the black hole surface gravity. This is the
fundamental difference with the standard case. Being the geometry
temperature dependent, it is not possible a priori to neglect
terms ordinarily linear in $\beta$: indeed, they are not in fact
linear in $\beta$. So, we cannot neglect the jacobian factor
$\Delta W$, because it is not simply proportional to $\beta$.
The correct approach then appears to be the
one of the quantum field theory partition function \zeqq, and, allowing for a
covariant choice of the functional measure, one could expect from
Thermofield Dynamics the same results than from ordinary finite
temperature quantum field theory.\par
A possible exception to the zeroth law ansatz is
represented by the extremal black hole:
a literal interpretation would give the value zero
for the temperature associated with the event horizon;
instead according to Hawking et al. point of view [\ross]
the temperature of the manifold is arbitrary and would
not be related to geometrical parameters.
Anyway, the very peculiar geometry of the extremal case
can justify its exceptionality, and the problem of the temperature
associated with an extremal black hole is still open.
A recent result due to Moretti [\moretti] is in the direction of
the impossibility to make finite temperature quantum field theory
on the extremal black hole manifold: from a QFT construction it results
that the only quantum state consistent with a quantum generalized
equivalence principle is the zero temperature one.\par
As a concluding remark we stress that the zeroth law ansatz, as
formulated in this work, does not necessarily clash against the idea
of black hole entropy as \ee, that still appears to be a major proposal
[\frono].\par
Further investigations are necessary in order to understand
full consistency of this approach [\work] and if it can give
some insights about the main problem of a statistical
interpretation of black hole entropy [\bhe].

\ACK{F.Belgiorno wishes to thank S.Liberati for his helpful
remarks.}

\vfill
\eject

\appendix

\centerline{1. Allen Theorem}

Our aim in this appendix
is first to generalize Allen's theorem from ultrastatic manifolds
to static ones, what we shall do, although not in fully rigorous form; and
then to discuss the result in light of the zeroth law ansatz.\par
We start reviewing Allen's distinction [\allen] between the
thermal partition function
$$
Z^{T}(\beta)=\sum_{n} e^{-\beta E_{n}}\equiv {\rm
thermal}\ Z
\eqn\zetter
$$
and the quantum partition function
$$
Z^{Q}(\beta)=\int_{P(\beta)} [D\Phi] e^{-S[\Phi]}\equiv {\rm
quantum}\ Z.
\eqn\zeqqal
$$
In \zeqqal\ $P(\beta)$ stays for the periodicity condition in
imaginary time (with period $\beta$) required for a scalar
field.\par
 In \zetter\ the index $n$ stays for all the multiparticle states
whose energy is $E_{n}$. For a free bosonic field \zetter\
becomes:
$$
Z^{T}(\beta)=\prod_{k} (1-e^{-\beta E_{k}})
\eqn\zetterbos
$$
where $k$ is the index for single particle states whose energy is
$E_{k}$. From \zetterbos\ it follows
$$
log Z^{T}(\beta)=-\sum_{k}log(1-e^{-\beta E_{k}})
\eqn\zetterdue
$$
that is the usual formula for bosonic particles.\pano
For quantum partition function \zeqqal\ the one-loop calculation is
exact and it can be expressed in terms of $\zeta$-function
[\hawzet]:
$$
logZ^{Q}(\beta)={1\over{2}}(\zeta^{\prime}(0,\beta)+\zeta(0,\beta)log({\mu}^2)).
\eqn\zeqqze
$$
$\mu$ is the renormalization scale usual in the $\zeta$-function
regularization.\pano
Allen shows that for an ultrastatic manifold (with compact spatial
sections) it holds
$$
logZ^{Q}(\beta)-logZ^{T}(\beta)=-{\beta
\over{2}}({d\over{dz}}+{1\over{2}} log{\mu}^{2})
z\zeta^{E}(z-1)|_{z=0}
\eqn\allenn
$$
where
$$\eqalign{
\zeta^{E}(2z)\Gamma (z)&=\int_{0}^{\infty} dt t^{z-1} Q(t)\cr
Q(t)&=\sum_{k} e^{-t E_{k}^{2}}.\cr}
\eqn\qze
$$
If $\zeta^{E}(z)$ is regular in $z=-1$ then the contribution in \zeqqal\
not present in \zetter\ is equivalent to a vacuum energy
contribution.
The difference between \zetter\ and \zeqqal\ does not affect
standard thermodynamics: it represents a shift independent from
the temperature for the free energy and the internal energy and
it does not modify the entropy.\pano
For the case of static manifolds it does not exist an analogous rigorous
demonstration about a difference between the two
partition functions. Anyway, a first natural extension of Allen
theorem is possible for conformally coupled scalar fields in a
static manifold: \logz\ allows to write
$$\eqalign{
logZ^{Q,conf}_{static}-logZ_{T}&=
logZ^{Q,conf}_{ultrastatic}-logZ_{T}\cr
&+logZ_{jacobian}.\cr}
\eqn\alconfal
$$
The jacobian is in fact the new contribution arising in the
static case.\par
{}From \logz\ and \alconfal\ the generalization to the nonconformal
coupling is
$$\eqalign{
logZ^{Q}_{static}-logZ_{T}&=
logZ^{Q}_{ultrastatic}-logZ_{T}\cr
&+logZ_{jacobian}+logZ_{non\ conformal\ jacobian}.\cr}
\eqn\jancal
$$
The dependence of the hamiltonian eigenvalues on $\beta$ in the case
in which Hawking effect is involved (they depend on the geometry)
and the use Allen does of a Mellin transform of the $\zeta$--function w.r.t.
$\beta$ in order to demonstrate \allenn\
don't allow to extend automatically Allen's theorem to manifolds
characterized by an intrinsic thermodynamics.

\vskip 2truecm
%\appendix2

\centerline{2. Thermofield Dynamics}

The formalism [\takume], [\lafl] is introduced in order
to express a statistical mean of an observable $A$ in terms
of a expectation value on a ``thermal vacuum state":
$$
<A>=Tr(\rho A)\equiv <O(\beta)|A|O(\beta)>.
\eqn\expe
$$
The equivalence in \expe\ is implementable only
in an extended Hilbert space.
Given a physical system described by an hamiltonian H in a
Hilbert space  ${\cal H}$, one has to introduce a fictitious
Hilbert space $\tilde{\calh}$ and an hamiltonian
$\tilde{H}$ describing an identical physical system.
Total Hilbert space and total hamiltonian are
$$\eqalign{
\calh_{tot}&=\calh \otimes {\tilde{\calh}}\cr
H_{tot}&=H-{\tilde{H}}.\cr}
\eqn\umz
$$
A physical (fictitious) n--particle state whose energy is $E_{n}$
is denoted by $|n>$ ($|\tilde{n}>$) and
$|n, \tilde{n}>=|n> |\tilde{n}>$ is the n--particle
tensor product state.\pano
Vacuum state in the extended Hilbert state is
$$
|O(\beta)>={1 \over {\sqrt{Z(\beta)}}} \sum_{n}\ e^{-\beta
E_{n}/2}\ |n,\tilde{n}>;
\eqn\vacfi
$$
$Z(\beta)$ is a normalization.\pano
In the calculation of \expe\ one has to
sum over all the fictitious states $\tilde{n}$; the result is
$$
<A>=<O(\beta)|A|O(\beta)>={1 \over {Z(\beta)}} \sum_{n}\ e^{-\beta
E_{n}}\ <n|A|n>.
\eqn\expa
$$
Between Thermofield Dynamics and black hole
thermodynamics, as discussed by Israel [\isr], the following
identifications are possible [\isr]
$$\eqalign{
|O(k)>&\equiv {\rm |HH-vacuum>}=|O(\beta)>\cr
<O(k)|A|O(k)>&=Tr_{II}(\rho A)\cr
&={1 \over {Z(k)}} \sum_{n}\ e^{-{{2 \pi}\over{k}}
E_{n}}\ <n|A|n>\cr
{\calh_{II}}&={\tilde{\calh}};\cr}
\eqn\isra
$$
$k$ is the surface gravity, and $|n>$ are n--particle (Boulware)
states.
Physical Hilbert space available to a static observer (region I)
is related to the physical Hilbert space of Thermofield Dynamics;
statistical means in \isra\ are relative to the region I, and the
fictitious space is identified with the space of states in region
II (time reversed of region I).\par
We note that the above correspondence was derived and in fact
holds ``on shell": the temperature $\beta$ is proportional to the
surface gravity $k$. \pano
If one allows the introduction of a conical defect in
treating Hawking radiation in Thermofield Dynamics, in the
sense that the mean w.r.t. HH--vacuum is taken but in the sum
$\sum_{n}\ e^{-{\beta} E_{n}}\ <n|A|n>$ the parameter
$\beta$ is off--shell, it happens that thermal equilibrium
condition is broken.\par

\refout
\bye